\documentclass{article}
\usepackage{dcase2019,amsmath,graphicx,url,times,booktabs, tabularx}

\usepackage{color, soul}
\usepackage{subcaption}
\usepackage{mwe}
\usepackage[hang,flushmargin]{footmisc}

\newcommand\blfootnote[1]{%
  \begingroup
  \renewcommand\thefootnote{}\footnote{#1}%
  \addtocounter{footnote}{-1}%
  \endgroup
}

\title{OtoMechanic: Auditory Automobile Diagnostics via Query-by-Example}

\twoauthors
  {Max Morrison}
    {Northwestern University\\
Computer Science Deptartment\\
     Evanston, IL, USA \\
     morrimax@u.northwestern.edu}
  {Bryan Pardo}
    {  Northwestern University \\
     Computer Science Department \\
     Evanston, IL, USA \\
     pardo@northwestern.edu}

\begin{document}

\ninept
\maketitle

\begin{sloppy}

\begin{abstract}
Early detection and repair of failing components in automobiles reduces the risk of vehicle failure in life-threatening situations. Many automobile components in need of repair produce characteristic sounds. For example, loose drive belts emit a high-pitched squeaking sound, and bad starter motors have a characteristic whirring or clicking noise. Often drivers can tell that the sound of their car is not normal, but may not be able to identify the cause. To mitigate this knowledge gap, we have developed OtoMechanic, a web application to detect and diagnose vehicle component issues from their corresponding sounds. It compares a user's recording of a problematic sound to a database of annotated sounds caused by failing automobile components. OtoMechanic returns the most similar sounds, and provides weblinks for more information on the diagnosis associated with each sound, along with an estimate of the similarity of each retrieved sound. In user studies, we find that OtoMechanic significantly increases diagnostic accuracy relative to a baseline accuracy of consumer performance. \blfootnote{This work was funded, in part, by USA National Science Foundation Award 1617497.}
\end{abstract}


\section{Introduction}
The timely maintenance of personal automobiles is vitally important to passenger safety. Foregoing important vehicle maintenance can cause a vehicle to behave unexpectedly and poses a danger to both occupants and nearby pedestrians. Failing to fix specific vehicle issues in a timely manner may also result in significantly more expensive repairs (e.g., engine damage due to a lack of oil). Because it is the consumer's decision to take their vehicle in for repair, it is of significant public interest to empower drivers with knowledge regarding the status of their vehicle, and whether any urgent repairs are needed.

The owner is typically alerted to a vehicle issue by either notifications from on-board computers or a change in the sensory experience of driving (e.g., a strange sound, smell, or vibration). On-Board Diagnostic (OBD) computer systems have been ubiquitous in consumer vehicles sold in the United States since 1996. While OBD systems provide some basic information about vehicle status directly to the driver (e.g., via the "Check Engine" light), a large majority of the diagnostic information from these systems must be retrieved via an external, specialized computer most consumers do not own.

Many automobile components in need of repair produce characteristic sounds. For example, loose drive belts emit a high-pitched squeaking sound, and bad starter motors have a characteristic whirring or clicking noise. Often drivers can tell that the car does not sound normal, but may not be able to identify the failing component. Consumer guides have been released to help drivers identify these sounds \cite{ftcrepair}. However, these guides assume that the consumer possesses significant knowledge, such as being able to locate and identify the vehicle components for power-steering or engine cooling. Many do not have this knowledge.

In this work, we present the OtoMechanic (``oto-'' meaning ``ear'') web application. OtoMechanic is designed for drivers who can hear a strange sound coming from their vehicle, but may be uncertain of the underlying issue and want information about it, such as the urgency and cost to repair. To use OtoMechanic, one uploads a recording of the sound a car is making and answers questions about when and where the sound happens. The system diagnoses the problem by measuring the similarity of the uploaded recording to reference recordings in a database of sounds produced by a variety of known vehicle issues. In the remainder of this article we describe related work, the methods used in OtoMechanic, the collection of labeled recording of automotive problems and a user study to evaluate the effectiveness of OtoMechanic. 

\begin{figure}
    \centering
    \begin{subfigure}[b]{0.475\textwidth}
        \centering
        \includegraphics[width=\textwidth]{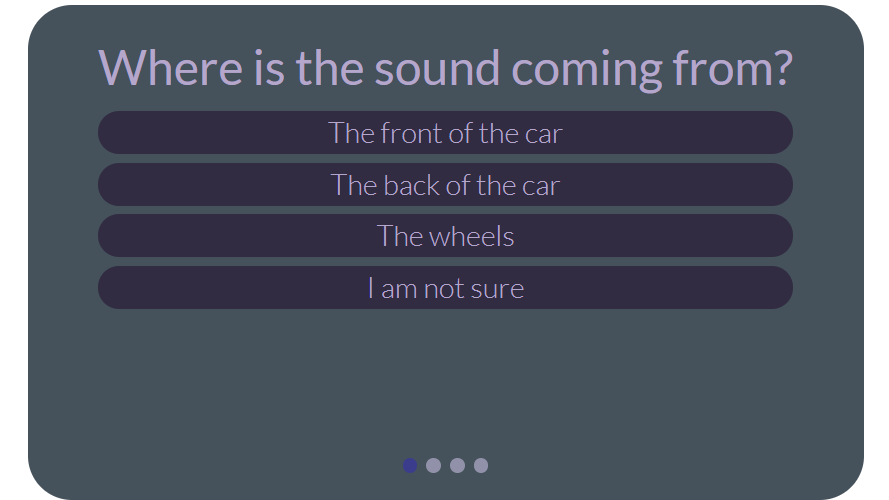}
        \subcaption%
        {{\small Specifying location information}}    
        \label{fig:where}
    \end{subfigure}
    \vskip\baselineskip
    \begin{subfigure}[b]{0.475\textwidth}  
        \centering 
        \includegraphics[width=\textwidth]{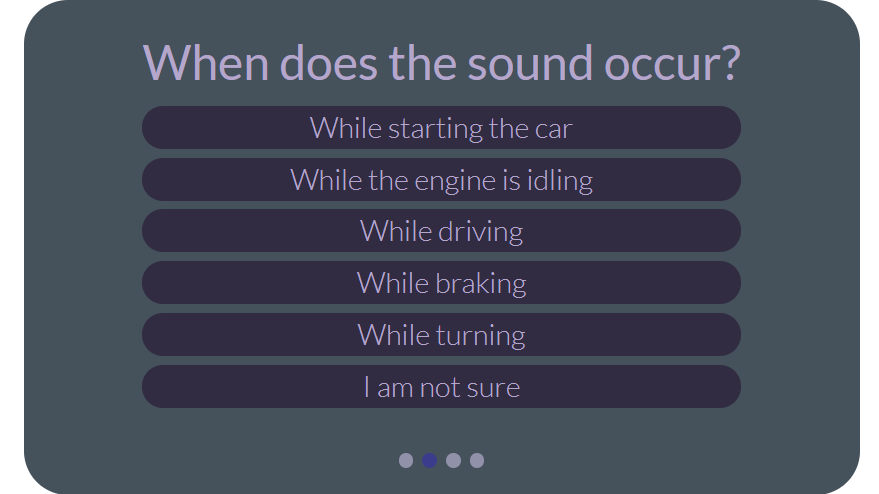}
        \subcaption%
        {{\small Specifying timing information}}    
        \label{fig:when}
    \end{subfigure}%
    \vskip\baselineskip
    \begin{subfigure}[b]{0.475\textwidth}   
        \centering 
        \includegraphics[width=\textwidth]{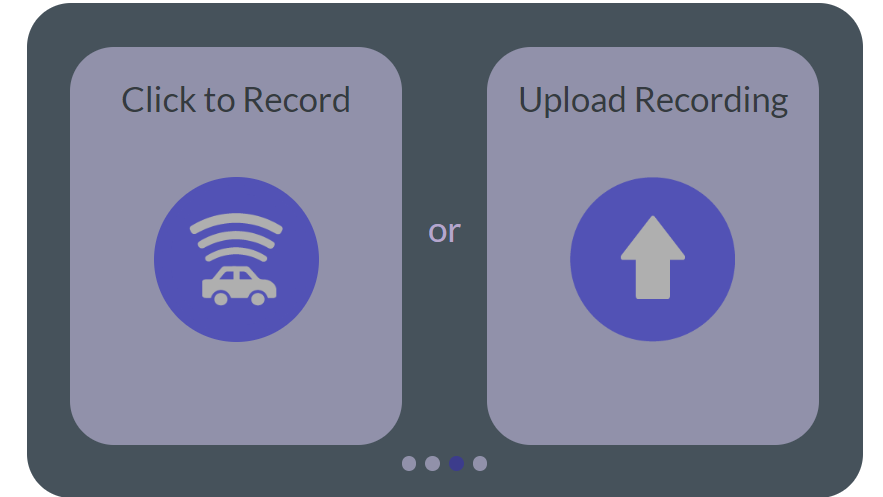}
        \subcaption%
        {{\small Audio upload interface}}    
        \label{fig:upload}
    \end{subfigure}
    \vskip\baselineskip
    \begin{subfigure}[b]{0.475\textwidth}   
        \centering 
        \includegraphics[width=\textwidth]{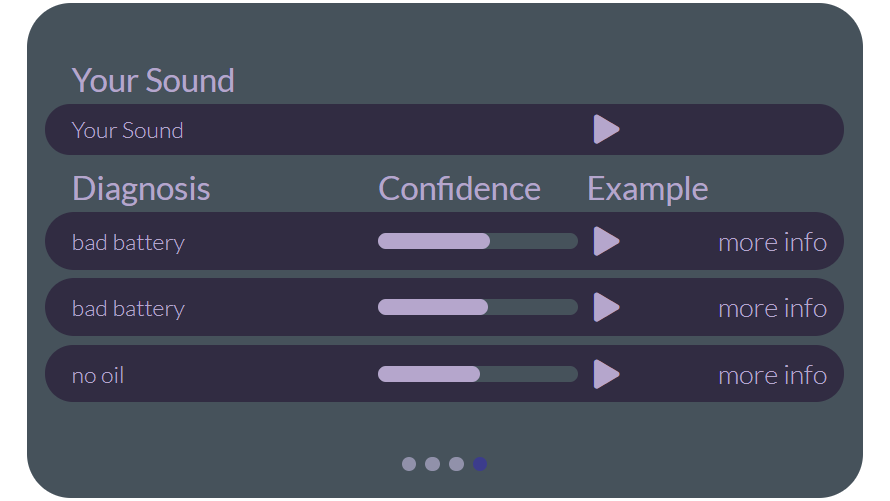}
        \subcaption%
        {{\small Diagnostic results interface}}    
        \label{fig:result}
    \end{subfigure}
    \caption[]
    {\small The OtoMechanic user interface }
    \label{fig:oto}
\end{figure}

\section{Related Work}
By far, the most reliable sources of diagnostic information are domain experts such as professional auto mechanics. However, the costs associated with visiting a mechanic cause consumers to hesitate and instead ask ``Do I need to go to a mechanic?'' Indeed, our user studies suggest that consumers often expend a significant amount of time in performing preliminary diagnostics before deciding to consult a mechanic. Since many drivers cannot identify specific causes or issues themselves, there have been efforts to build software to perform diagnosis of automotive issues

Recent work by Siegel et al. \cite{siegel2018automotive} demonstrates that convolutional neural networks achieve a 78.5\% accuracy rate for visually diagnosing photos of damaged and unsafe tires---a significant improvement over the 55\% accuracy achieved by humans not trained to detect flaws in tires. This shows the potential for systems to diagnose vehicle problems with an accuracy that exceeds non-expert performance. While Siegel et al. diagnose photos of tires (i.e., visual information), our system diagnoses the sounds produced by the vehicle.

A strange sound is a useful indicator for a variety of specific vehicle issues (e.g., worn drive belts and low battery). Unfortunately, many people cannot name which component is failing from the sound. Datasets of annotated car sounds offer one resource for consumers trying to identify a component making a strange sound. Existing datasets include the Car Noise Emporium \cite{cartalk} consisting of vocal imitations of 45 vehicle issues, the ClingClanger mobile application consisting of 27 actual recordings of vehicle issues taken from a single vehicle \cite{mcmn}, and YouTube. YouTube videos demonstrating the sounds of common car issues have millions of views \cite{wheelbearing, swaybar}. However, finding a video to diagnose a specific issue requires consumers to identify by name the failing component in order to construct a useful search query.  

In all of these approaches (including ClingClanger), the task of matching the sound emitted by their vehicle to example sounds with known causes falls to the user, who is required to listen to all of the available recordings to find the best match. The amount of audio that must be listened to increases linearly as the number of possible diagnostic sounds increases, imposing a significant time cost.

Auditory analysis is regularly used by professional test engineers and mechanics when diagnosing vehicle issues, and a large body of literature exists on this subject, specifically for engine diagnostics \cite{delvecchio2018vibro, deptula2016acoustic, figlus2016use, henriquez2014review, navea2013design, ning2016auto, siegel2016engine}. The work most similar to ours is by Nave and Sybingco \cite{navea2013design}, who perform a classification task on the sounds of three engine issues, and consider the high variance of sounds caused by the same diagnosis across different vehicles. Nave and Sybingco developed an application for the Android operating system to perform classification of these three engine sounds. However, this app is only useful when the issue has already been narrowed to an engine fault, which is itself a non-trivial diagnostic task. More generally, all of these works focus on the development of tools for professional use, and do not address use by non-expert consumers.

\section{OtoMechanic}
OtoMechanic is an end-user application for diagnosing automotive vehicle issues from an audio recording. It can diagnose many more issues than prior systems and is designed to be accessible by people with little or no expertise in automotive diagnosis or repair. 

OtoMechanic asks users to provide two inputs: 1) a recording of a troubling sound coming from the user's vehicle and 2) answers to the questions: "Where and when does the sound occur?" Given this information, it queries a database of sounds associated with vehicle issues (see Section \ref{otomobile}), narrowing the search based on when and where the query sound occurs. Users are then presented with the top 3 matching sounds, ordered by similarity with the user's recording, as well as the diagnoses and confidence level for each matching sound. For each retrieved sound, web links with more information about the diagnosis are provided. This helps users conduct further research on their vehicle's issue.

\subsection{Interface Design}
The OtoMechanic interface (Figure \ref{fig:oto}) presents, in sequence, four distinct displays to the user. The first display (Figure \ref{fig:where}) asks where on the vehicle the concerning sound is being produced. Users may respond with  \textit{front}, \textit{rear}, or \textit{wheels} of the vehicle, or indicate that they are not sure. The second display (Figure \ref{fig:when}) asks users when the concerning sound occurs. Users may respond with \textit{while starting the car}, \textit{while the engine is idling}, \textit{while driving}, \textit{while braking}, or \textit{while turning}, or indicate that they are not sure. After specifying when and where the sound occurs, users are prompted to upload a recording of the sound (see Figure \ref{fig:upload}).

Figure \ref{fig:result} shows example diagnostic results from OtoMechanic. In this example, a recording of a failing battery was uploaded, and no timing or location information was given. The diagnostic results allow users to listen to their uploaded recording and compare it to the three most relevant matches, as determined by our system. The diagnosis corresponding to each returned audio file is provided to the user, as well as a visual indicator of the confidence of the diagnosis and a weblink to Google search using curated search terms relevant to that issue. The search terms used ensure that both text and video descriptions of the diagnosis are included within the first page of search results. This allows users to verify the accuracy of the diagnosis, and determine if they need to take further action (e.g., taking their car to a mechanic for repair).

\subsection{Diagnostic Procedure}
Once the user provides their audio recording and any qualitative input (i.e., where and when the sound occurs), our system queries a database of annotated sounds of vehicle issues to determine the most likely diagnosis. To do this, our system first filters out all audio files in the database that are inconsistent with the time and location information provided by the user. Our system then computes the similarity between the user's audio recording and each audio file in this list of potential matches.

To compute the similarity between the user's audio recording and a recording in our database, we split each audio file into equal-sized slices of 960 milliseconds at a sampling rate of 16,000 Hz. Each 960 millisecond slice is transformed by passing it through a pretrained VGGish neural network \cite{vggish}. This network is trained to classify audio taken from millions of YouTube videos. We make use of the modification proposed by Kim and Pardo \cite{kim-qbv}, who show that a feature vector formed from the output of two neural network layers of the trained VGGish model is significantly better for a query-by-example task than using only the output of the final layer.

The final feature vector representation of a single audio recording is the element-wise average of the VGGish feature vectors extracted from all 960 ms slices from that recording. The similarity of a user's recording to a recording in our database is the cosine similarity between these fixed-length feature vectors. In informal experiments on both a commodity laptop and an Amazon Web Services EC2 t2-micro instance, inferring the most relevant audio recordings in the OtoMobile dataset of automotive sounds (see Section \ref{otomobile}) takes between 200-500 milliseconds for user recordings up to 10 seconds in length, making it suitable for interactive use.

For each retrieved recording, we report a confidence score to the user. Confidence scores are derived by mean-shifting the similarity scores to 0.5 and scaling them to range between 0 and 1 using the mean and range computed over all pairwise similarity scores on the OtoMobile dataset (excluding self-similarities). To minimize the need for users to listen to many recordings, only the top three most similar recordings are displayed.

Currently, OtoMechanic selects one of 12 possible diagnoses. The number of possible diagnoses is determined by the number of different diagnoses that have representation in the dataset of vehicle sounds. As sounds relevant to new vehicle issues are placed in the dataset, OtoMechanic becomes able to suggest these new issues as possible diagnoses.  

\subsection{OtoMobile Dataset} \label{otomobile}

As no large collection of audio recordings of vehicle issues existed, we curated our own dataset, called the OtoMobile dataset. OtoMobile consists of 65 recordings of vehicles with failing components, along with annotations. These annotations include the diagnosis of the failing component (one of 12 common automobile issues), the location of the component on the car (\textit{front}, \textit{rear}, or \textit{wheels}), the time during which the sound occurred (\textit{while starting}, \textit{while idling}, \textit{while driving}, \textit{while braking}, or \textit{while turning}) the video name and URL, and the start location of the video where the sound was extracted. Excerpts were selected based on the following criteria:

\begin{itemize}
    \item The diagnosis of the sound coming from the vehicle was provided by either a professional auto mechanic, or someone who had consulted a professional auto mechanic to diagnose the sound.
    \item At least one second of audio of the problematic vehicle sound was available, during which other noises (e.g., speech) were absent.
    \item No more than one recording of the same diagnosis was extracted from each video.
\end{itemize}

\noindent
The selected audio recordings were cropped to contain only the problematic sounds and normalized to all have the same maximum amplitude. The dataset is available for download for educational and research purposes\footnote{https://zenodo.org/record/3382945\#.XXCDG-hKhPY}. Despite the availability of the OtoMobile dataset, we note that data scarcity is still a major bottleneck in auditory vehicle diagnosis.

\section{User Studies}
We hypothesize that our system improves the ability of nonexperts to identify the vehicular problem causing a particular sound. To test our hypothesis, we conducted two user studies. The goals of the first study were two-fold: 1) understand the existing methods that consumers use to diagnose strange noises coming from their vehicles and 2) determine a baseline accuracy for non-expert diagnosis of such noises. The goal of the second study was to determine whether OtoMechanic can be used to improve non-expert auditory vehicle diagnosis relative to the baseline accuracy determined in our first study.

Both studies made use of Amazon Mechanical Turk (AMT) to recruit and pay participants. We required participants to be from the United States and have at least a 97\% acceptance rate on AMT to qualify for our study. 86 participants completed the diagnostic baseline study study and 100 participants completed the system evaluation study. 

\subsection{Establishing a Diagnostic Baseline} \label{sec:prelim}
The purpose of our first user study was to understand the existing methods people use to diagnose troublesome vehicle sounds and the efficacy of their methods. In this study, each of our participants was presented a randomly selected audio recording from the OtoMobile dataset (see Section \ref{otomobile}), along with information about when (i.e. "when turning", "when idling") and where (i.e. "from the wheels") the troublesome sound occurs. Participants were asked to write a brief description of the steps they would take to diagnose the vehicle, given this information.

After participants described their approach, a new portion of the questionnaire was revealed to them, where we asked them to actually diagnose the sound they had been presented in the previous question. Participants were presented a 12-way forced choice selection of diagnoses. Note that all sounds in the OtoMobile dataset are due to one of 12 issues and that the data set is balanced by issue, so selecting an answer at random will be correct 1/12 of the time. 

\subsection{Diagnostic Baseline User Study Results}
Participant descriptions of their diagnosis method indicated a strong preference for manual inspection as a diagnostic method. Of the 86 diagnostic methods described by participants, 73 of the descriptions mentioned a physical interaction with the vehicle that they would use to gain more information on the issue. We found that participants were far more willing to actively participate in diagnosing the vehicle than to seek out a professional auto mechanic; only 23 of our 86 participants mentioned interacting with a mechanic in order to diagnose their vehicle.

Despite the wide availability of online resources, our results suggest that participants are unlikely to connect with these resources. Only four participants mentioned they would use online resources to assist their diagnosis. We hypothesize that this may be due to the difficulty that non-experts face in constructing relevant search terms. For example, one participant describes a process of first isolating symptoms of the vehicle and then using that information to construct a search query: ``I'd see what factors might affect [the sound], e.g. stepping on the gas, changing gears. Then I'd look up potential answers on the Web.'' The ability for OtoMechanic to connect users to online resources by providing relevant web links associated with a diagnosis represents a potential solution to the lack of utilization of online resources that we observed.

Only two participants mentioned asking a friend or family member, and only two participants mentioned accessing their car's computer via an external OBD reader. Responses indicate a large variance in experience with vehicle repairs. One participant self-reported that they ``don't know anything about cars'', while others indicated significant knowledge of car components and prior experience replacing brake pads, fuel pumps, and drive belts.  

When asked to select the diagnosis that best corresponded to a randomly selected audio recording from the OtoMobile dataset, participants were able to identify the correct diagnosis with 37.2\% accuracy. Chance performance on this 12-way forced choice task is 8.3\%.

\subsection{System Evaluation User Study} \label{sec:oto-exp}
We evaluate the efficacy of the OtoMechanic approach through a second user study. This study was performed by a new group of 100 participants. Each study participant was given a random audio recording from the OtoMobile dataset and the same qualitative information as in the first study (i.e., when and where the sound occurred). As with the diagnostic portion of the previous study, participants were presented a 12-way forced choice selection of diagnoses and chance performance on this task was 1/12. 

In this study, participants were asked to use OtoMechanic to diagnose the audio recording, and provide their diagnosis by selecting from the same diagnostic options as in our first user study. To prevent trivial similarities, the audio recording provided to each participant was removed from the list of possible matches retrieved by OtoMechanic.

\begin{table}[!htb] 
  \centering
    \begin{tabular}{l|r}
        \textbf{Diagnostic Method} & \textbf{Accuracy (\%)}\\
        \hline
        Random & 8.3\\
        Human Baseline & 37.2\\
        OtoMechanic (no time or location information) & 34.8\\
        Random (with time and location information) & 39.3\\
        Humans using OtoMechanic & 57.0\\
        Oracle System Performance & 58.7\\
    \end{tabular}
    \caption{Diagnostic accuracies of experiments on the OtoMobile dataset} \label{tab:dem-gender}
\end{table}

\subsection{Oracle System Classification Results}
We present the diagnostic accuracy of selecting the diagnosis corresponding to the best matching audio recording determined by OtoMechanic on a 12-way classification of the OtoMobile dataset. When the location and time information is included as input alongside the audio recording, our system achieves an accuracy of 58.7\% on the OtoMobile dataset. The likelihood of the correct diagnosis being one of the 3 results returned to the user by OtoMechanic (i.e., the top-3 accuracy) is 82.9\%. Without using the information about when and where the sound occurred, our system achieves an accuracy of 34.8\%, with a top-3 accuracy of 53.0\%. If we use the location and time information but choose randomly from the recordings matching those criteria, our system achieves an accuracy of 39.3\%. This indicates that the knowing when and were the sound occurs is a significant factor in diagnostic accuracy.

\subsection{System Evaluation User Study Results}
When asked to use OtoMechanic to diagnose a troublesome vehicle sound, our 100 participants achieved a diagnostic accuracy of 57.0\%. This is nearly equal to the 58.7\% accuracy achieved by simply selecting the top choice indicated by OtoMechanic and significantly greater than the 37.2\% achieved by participants on the identical task when performed without access to OtoMechanic. This indicates that our application is significantly more effective at diagnosing vehicle issues than a baseline of prior knowledge of participants.

\section{Discussion}
Vehicle failure poses serious risk to public health. Road traffic crashes are the 9th leading cause of death worldwide, and in the US alone there are an estimated 800 deaths and 47,000 injuries each year due to vehicle failures \cite{nhtsa, asirt}. OtoMechanic provides a convenient method for informing consumers of potential failures due to damaged vehicle components. Our user studies showed use of OtoMechanic significantly increases  the accuracy of vehicle diagnosis from troublesome sounds without requiring expert knowledge of vehicle maintenance or repair. Our application connects non-expert consumers to resources that enable them to make educated decisions about their vehicle. In future work, we aim to increase the size of the OtoMobile dataset and the specificity of the annotations (e.g., the make and model of the vehicle). We also aim to compare our system against more challenging baselines (e.g., the performance of professional auto mechanics) and investigate the utility of OtoMechanic in on-board vehicle computers.

\bibliographystyle{IEEEtran}
\bibliography{refs}

\end{sloppy}
\end{document}